\documentstyle[aps,pre,graphicx,multicol,amssymb,amsmath]{revtex}

\draft
\tighten

\newcommand{\smol}{Smoluchowski}
\newcommand{\smoleq}{Smoluchowski's equation}
\newcommand{\avecs}{S}

\newcommand{\Cpar}{C_{\rm D}} 
\newcommand{\Cf}{C_v} 

\begin{document}

\title{Dynamic Scaling in One-Dimensional 
Cluster--Cluster Aggregation}
\vspace{0.8cm}
\author {E.~K.~O.~Hell\'{e}n$^1$, T.~P.~Simula$^1$, and
M.~J.~Alava$^{1,2}$} 
\address{$^1$ Helsinki University of Technology, Laboratory of
Physics, P.O.Box 1100, FIN-02015 HUT, Finland}
\address{$^2$ NORDITA, Blegdamsvej 17, DK-2100 Copenhagen, Denmark.}

\date{\today}

\maketitle

\begin{abstract}

We study the dynamic scaling properties of an aggregation model in which
particles obey both diffusive and driven ballistic dynamics. The
diffusion constant and the velocity of a cluster of size~$s$ follow
$D(s) \sim s^\gamma$ and $v(s) \sim s^\delta$, respectively. We
determine the dynamic exponent and the phase diagram for the
asymptotic aggregation behavior in one dimension in the presence of 
mixed dynamics. The asymptotic dynamics is dominated by the
process that has the largest dynamic exponent with a crossover that is
located at $\delta = \gamma - 1$. The cluster size
distributions scale similarly in all cases but the scaling function
depends continuously on $\gamma$ and $\delta$. For the purely
diffusive case the scaling function has a transition from exponential
to algebraic behavior at small argument values as $\gamma$ changes
sign whereas in the drift 
dominated case the scaling function decays always exponentially.

\end{abstract}


\begin{multicols}{2}[]

\section{Introduction}

Both reaction-- and diffusion--limited cluster--cluster
aggregation~(DLCA) have been successfully used to understand the
dynamics of colloidal
aggregation~\cite{Meakin:PhysicaScripta46}. These models predict well  
both the structure of aggregates and the growth behavior in dilute
particle suspensions as long as the dynamics is 
dominated by Brownian diffusion. As the growth of the aggregates
proceeds the 
sedimentation of clusters due to gravitation becomes more
pronounced altering the growth mechanism and cluster structure. This
was recently observed in experiments~\cite{Allain:JCIS178}. 

The purpose of this paper is to study dynamic scaling 
in one-dimensional cluster--cluster aggregation
in the presence of a competition between diffusion and drift.
We show that the dynamics at long times is dominated by the
aggregation process which by itself would lead to the fastest
growth. The conventional mean-field
theory gives the correct dynamic exponent for the
field dominated case but fails when diffusion dominates. 
The mean-field theory also predicts that the scaling function of the
cluster size distribution in the diffusive (driven) case would
drastically change when $\gamma$ ($\delta$) changes sign. 
Such a transition is observed for the diffusive case but not for the
driven one. The dynamic phase diagram 
shows four different regions depending on the relative rates of the
diffusion and drift. 

The paper is organized as follows. Section~\ref{modelsec} introduces
the model and describes the algorithm used in simulations. 
In Section~\ref{analsec} the dynamic scaling is studied using
mean-field rate equations approach. The mean-field results are
compared to simulations in section~\ref{simsec}. Section~\ref{dissec}
concludes the paper with a discussion. 

\section{Model} \label{modelsec}

The field--driven cluster--cluster aggregation~(FDCA) model is defined
on a one-dimensional lattice with periodic 
boundary conditions, for simplicity. Initially particles are
distributed randomly on a lattice of $L$ sites up to a
concentration~$\phi$. Sites connected via nearest neighbor occupancy
are identified as belonging to the same cluster. 
The diffusion coefficient of a cluster of size $s$ takes the
form $D(s) = D_1 s^\gamma$ where $\gamma$ is the diffusion exponent
and $D_1$ a non-negative constant. 
The clusters are also driven into one direction with a size dependent
drift velocity $v(s) = v_1 s^\delta$, which defines the field
exponent~$\delta$.

In simulations a cluster is selected randomly and the time is incremented
by $N(t)^{-1}\Omega_{\rm max}^{-1}$, where $N(t)$  
is the number of clusters at 
time~$t$ and $\Omega_{\rm max}$ is the maximum mobility of any of the
clusters in the system at that time. 
The cluster mobility is defined as 
$\Omega(s) =  \Cf s^\delta + 2\Cpar s^\gamma$
where  $\Cf$ and $\Cpar$ are non-negative constants.
The choice $\Cf=0$ gives normal DLCA. The cluster is moved 
only if $x<\Omega(s)/\Omega_{\rm max}$, where $x$ is an
uniformly distributed random number in the interval~$[0,1]$. The step
is taken along (against) the field 
with probability $p$ ($q$), where $p=(\Cf s^\delta+\Cpar
s^\gamma)/\Omega(s)$ and $q=1-p$. If
after the move two clusters are in contact, they are
irreversibly aggregated together. Note that time is increased
for each attempted move.

\begin{figure}
\centering
\includegraphics[angle=-90,width=0.95\linewidth]{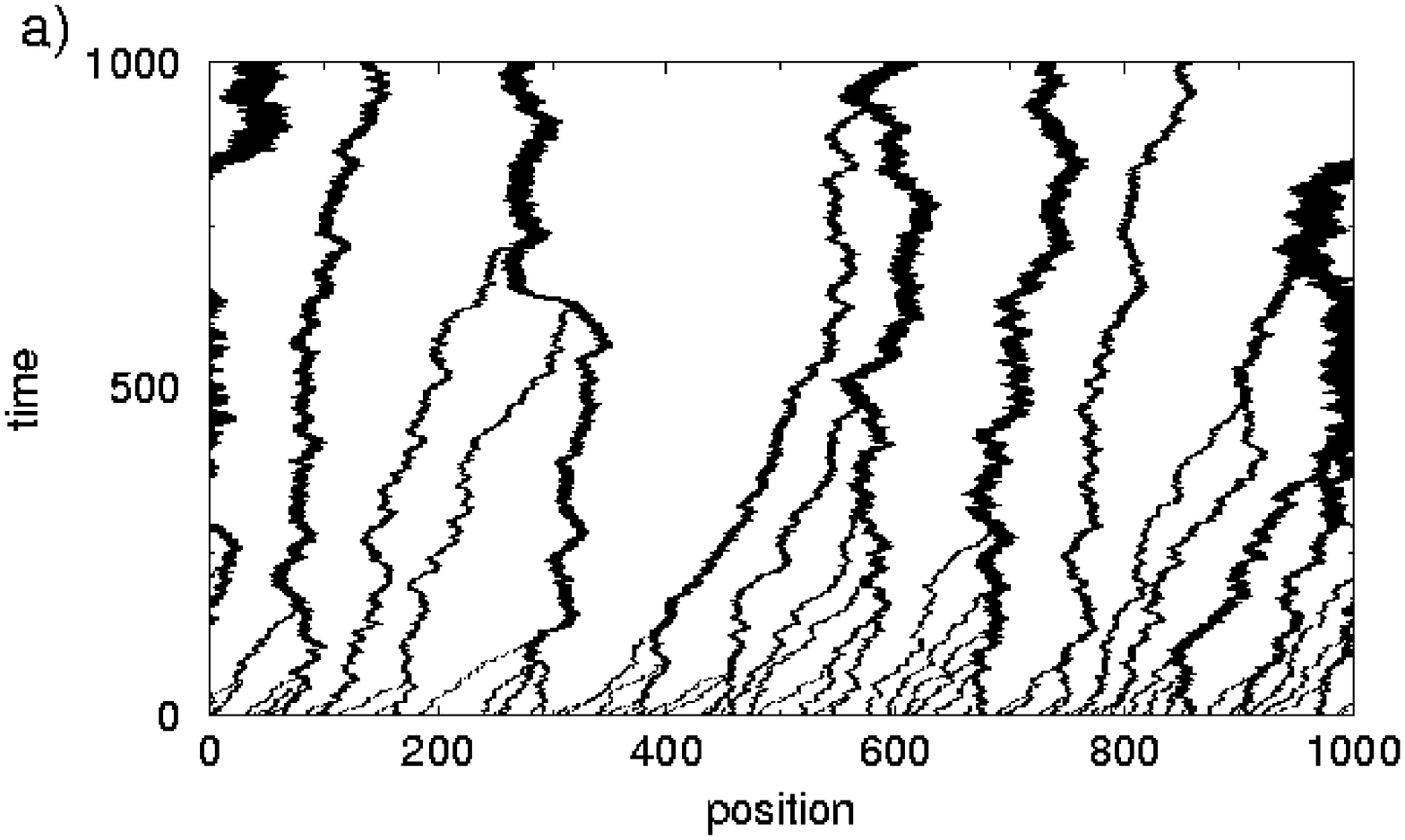}
\includegraphics[angle=-90,width=0.95\linewidth]{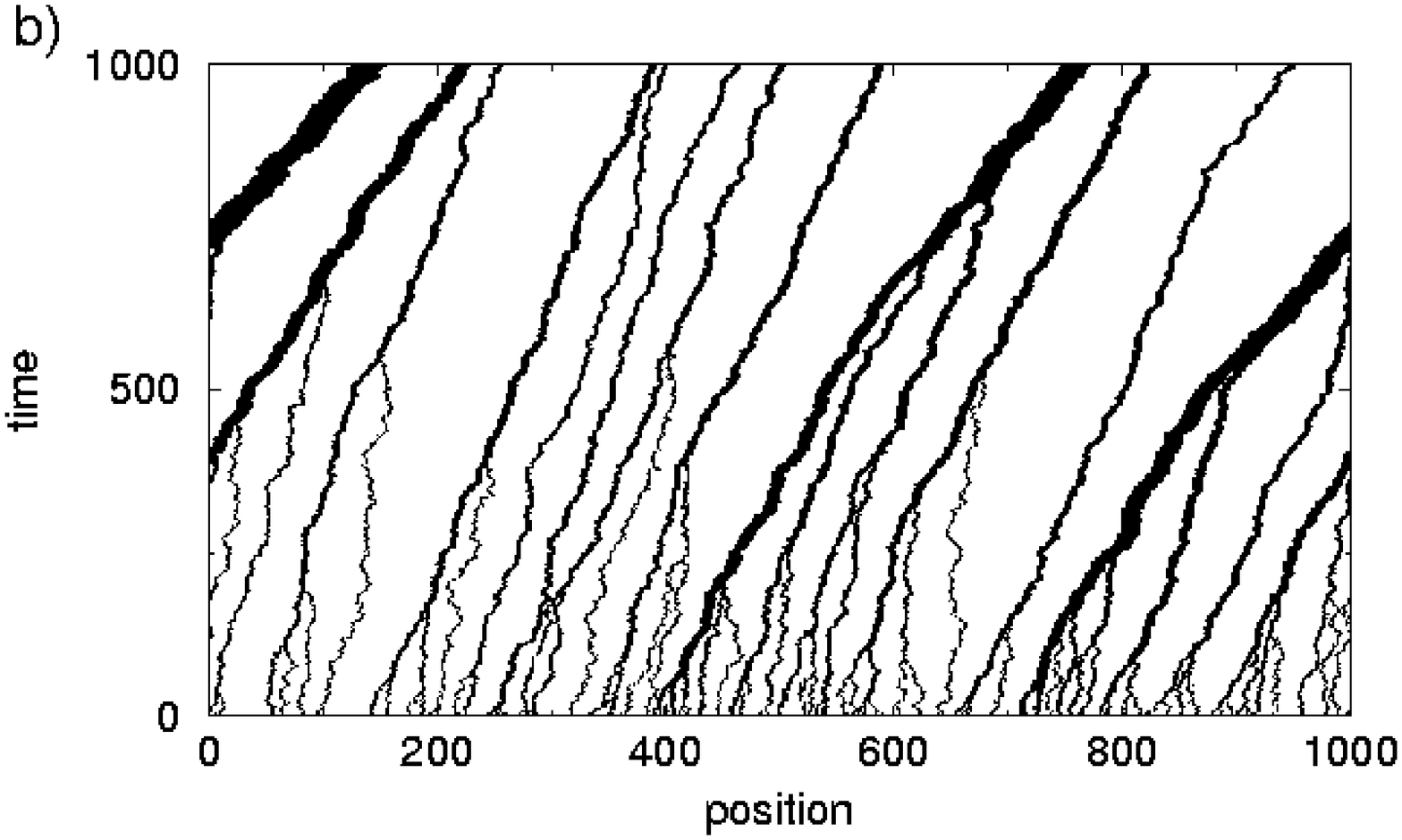}
\caption{An example of the dynamics in FDCA for $\phi = 0.1$,
a) $\gamma = 0.5$, $\delta = -1.0$ and
b) $\gamma = -1.0$, $\delta = 0.5$. System size $L=1000$.
The time scales are normalized differently.
} \label{spacetimeplot} 
\end{figure}

Figure~\ref{spacetimeplot} shows an example of the dynamics when
either the diffusion (fig. 1a) or the drift (fig. 1~b) dominates the 
large time aggregation behavior. The diffusion and field exponents are
chosen in such a way that at large times the largest clusters are the
most mobile ones. In the latter subfigure, notice the clear
breaking of the reflection symmetry in the cluster dynamics as the
drift begins to dominate. Similar
behavior is visible in the early-time dynamics of the diffusion-dominated
case.

\narrowtext

\section{Scaling Analysis} \label{analsec}

Before considering any specific aggregation rules let us first
represent the well-known mean-field approach. We want to
compare different dynamical processes in order to find the dominating
aggregation mechanisms. 
Denote the number of clusters of size $s$ per site at time $t$ by
$n_s(t)$ and the mean cluster size by $\avecs(t)$. 
The mean-field description of irreversible aggregation which neglects
spatial correlations is given by
\smoleq~\cite{MvonSmoluchowski:PhysZ17_593_1916} 
\begin{equation}
\frac{{\rm d}n_s}{{\rm d}t} =  \frac{1}{2}
\sum_{i+j=s}
K(i,j) n_{i} n_{j} - \sum_{i=1}^{\infty} K(i,s)
n_{i} n_s, \label{Smoluchowski}
\end{equation}
where the reaction kernel $K(i,j)$ describes the rate at which
clusters of size $i$ and $j$ aggregate. It is assumed
to be a homogeneous function $K(ai,aj) = a^\lambda K(i,j)$ with
$K(i,j) \sim i^\mu j^{\lambda-\mu}$ for $i \ll j$. Kernels are
classified by $\mu$~\cite{vanDongen:PRL54}: $\mu>0$ (class~I), 
$\mu=0$ (class~II), and $\mu<0$ (class~III). Independent of the class
the solution scales for mass conserving systems as $n_s(t) =
\avecs(t)^{-2}f(s/\avecs(t))$. In 
class~I the aggregation is dominated by the collisions of large
clusters with large ones whereas the dominant contribution in
class~III comes from the reactions between large and small
clusters. In class~II these two processes are equally important. 
The class~III processes can be identified from the form
of the scaling function since in classes~I and II $f(x) \sim
x^{-\tau}$ but in class~III $f(x) \sim \exp(-x^{-|\mu|})$ as 
$x \to 0$~\cite{vanDongen:PRL54}. 

Here we concentrate on the scaling function, on the
polydispersity exponent $\tau$, and 
on the dynamic exponent $z$
describing the growth of the mean cluster size: $\avecs(t) \sim t^z$.
The polydispersity exponent in the mean-field~(\textsc{mf}) is
easily found to be $\tau_{_{\rm MF}}=1+\lambda$ 
in class~I. Predicting it for class~II processes is still a
challenge~\cite{Cueille:PRE55}.
However, for all non-gelling systems, {\it i.e.} $\lambda \le 1$,
the dynamic exponent
is related to the homogeneity exponent~$\lambda$ as 
$z_{_{\rm MF}}=1/(1-\lambda)$~\cite{vanDongen:PRL54}.

The upper critical dimension, above which the mean-field theory is
exact, may be calculated once the reaction
kernel is known~\cite{vanDongen:PRL63}. 
Consider for a moment the aggregation of clusters of fractal dimension
$d_f$ in $d$ dimensions. 
For a DLCA kernel $K_{\rm D}(i,j) \sim (i^{1/d_f} + j^{1/d_f})^{d-2}
(i^\gamma+j^\gamma)$ ($d \ge 2$) the mean-field theory is not exact in
any finite 
dimension~\cite{vanDongen:PRL63} but the deviations are negligible
already in $d=3$~\cite{Ziff:JCP82}. 
In the driven case, if diffusion and velocity 
fluctuations are neglected, clusters move
ballistically. The collision   
probability of two clusters is proportional to the product of the
mutual cross-section of the clusters and the velocity difference
between clusters $K_v(i,j) \sim (i^{1/d_f}+j^{1/d_f})^{d-1}
|i^\delta-j^\delta|$. Thus in the mean field description the driven system in
$d$ dimensions has similar scaling properties as the 
diffusive one in $d+1$ dimensions and therefore the upper critical
dimension is infinite for both.

If both diffusion and drift are
present the faster dynamics, as measured by the 
associated dynamic exponent, could be expected to dominate. This is
verified by the simulation results, discussed in the next section. 
Thus it is adequate to consider the two
dynamic processes separately. For example, in one dimension the
scaling properties of $K_v$ necessitate that $\lambda = \delta$
together with $\mu = \delta$ for $\delta < 0$ (class~III) and $\mu =
0$ for $\delta \ge 0$ (class~II). Thus the scaling function should
drastically change as $\delta$ changes its sign. 
In one dimension the collision cross-section   
is independent of the clusters' sizes. Thus the above
scaling analysis is directly applicable to the diffusion--limited
case, too and there should be similar transition between the
classes~III and II at $\gamma=0$.

In one dimension the scaling properties of the reaction kernels
together with $z_{_{\rm 
MF}}=1/(1-\lambda)$ give the mean-field dynamic exponent in the
diffusive and driven cases as $z_{_{\rm MF}}=1/(1-\gamma)$ and $z_{_{\rm
MF}}=1/(1-\delta)$, respectively. The strong fluctuations
are responsible for the fact that the correct exponent is
 $z=1/(2-\gamma)$ in the
diffusive case~\cite{Kang:PRA33,Miyazima:PRA36}. The dynamic
exponent may on the other hand be obtained more simply 
by considering the two length scales coming from 
the two dynamical processes: the diffusive length scale $l_D \sim
\sqrt{Dt}$ and the ballistic one $l_v \sim vt$. Naturally the average
cluster size is proportional to the dominant length scale, {\it
i.e.} $S(t) \sim l$, which together with 
$D(s) \sim s^\gamma$ and $v(s) \sim s^\delta$
results in $z=1/(2-\gamma)$ and
$z=1/(1-\delta)$ for the diffusion and drift dominated cases,
respectively. The simulation results presented in section~\ref{simsec}
confirm these arguments. 
Thus the \smol\ approach predicts the correct dynamic exponent for the
driven case even in one dimension.  
If both diffusion and drift are present 
$z=\max \{1/(2-\gamma),1/(1-\delta)\}$ with the crossover at 
$\delta = \gamma-1$.

The average cluster size at the crossover can be estimated by comparing 
the pairing time (the time required for
$\avecs \rightarrow 2 \avecs$) due to diffusion, $t_{\rm agg}^D$, to
that due to drift, $t_{\rm agg}^v$. In the diffusive case the
pairing time can be obtained by considering a random walk on
coarse-grained system with the lattice constant set equal to the
average cluster radius $R$~\cite{Kolb:PRL53}. In one dimension the
cluster density on the lattice is 
$
\rho(t) = N(t)/V = \phi,
\label{clusterdensityeq} 
$
where the volume $V=L/R$.
A cluster travels a distance of its own radius diffusively in time
$R^2/D$. As it takes on the average $\rho^{-2}$
steps to pair up $t_{\rm agg}^D = R^2/(D\rho^2)$.

For driven clusters the variation in cluster
velocities is the relevant parameter. Therefore the pairing time is of
order $t_{\rm agg}^v = R/(\sigma_v\rho)$, where $\sigma_v =
\sqrt{\langle v^2 \rangle - \langle v \rangle^2}$ is the standard
deviation of the cluster velocities. It can be
calculated from the velocity distribution $p(v) = sn_s \left|
\partial s(v)/\partial v \right|$ which gives
$
\sigma_v \approx v_1 S^\delta \sqrt{I_2-I_1^2},
$
where $I_\alpha = \int d{\rm x} x^{\delta \alpha +1} f(x)$ and the 
approximation comes from replacing the sum by an integral. 
The proportionality constant $A=\sqrt{I_2-I_1^2}$
has to be determined numerically from simulations since calculating it
would require the knowledge of the whole scaling function.
The crossover takes place as $t_{\rm agg}^v
\approx t_{\rm agg}^D$ which gives the average cluster size at the
crossover as
\begin{equation}
S_{\rm cross} \approx \left( \frac{2D_1\phi}{Av_1r_0} 
\right)^{1/(\delta-\gamma+1)} \label{Scross1d}
\end{equation} 
where $r_0$ is the elementary particle radius.

\section{Simulations} \label{simsec}

In simulations the system sizes range from $5\times10^5$ to
$2\times10^6$, the data are averaged over $50-2000$ realizations, the
concentration is usually at $\phi=0.1$, and random initial conditions
are used. Neither the 
initial conditions nor the concentration has
any effect on the asymptotic dynamic scaling properties as was
verified by simulations.  
The timescale is fixed by
setting $\Cpar=1$ for DLCA and $\Cf=1$ for FDCA if not otherwise
mentioned. The 
mean cluster size is calculated using both the number ($k=1$)
and weight averages $(k=2)$
\begin{equation}
\avecs_k(t) = \sum_{s=1}^{\infty} s^kn_s(t) \left/ \sum_{s=1}^{\infty}
s^{k-1}n_s(t)\right. .
\end{equation}
Both averages scale similarly and the number average is used in all the
figures following. In order to ensure that the scaling regime is
reached the dynamic exponent is calculated using the method of
consecutive slopes~\cite{Barabasi}. 

\narrowtext
\begin{figure}
\centering
\includegraphics[angle=-90,width=\linewidth]{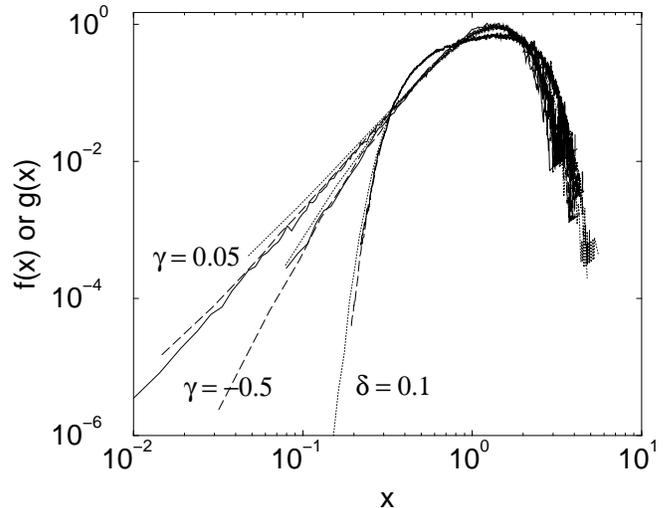}
\caption{The numerically obtained scaling functions as a function of
the scaling variable 
$x=s/\avecs(t)$ for DLCA ($\gamma$) and FDCA ($\delta$) at times
$10^4$~($\cdots$), 
$10^5$(-- --), and $9 \cdot 10^5$~(---). System sizes and number of
realizations are $(5\times10^5$, $50)$, $(5\times10^5$, $1000)$, and
$(2\times10^6$, $2000)$ for $\gamma = -0.5$, $\gamma = 0.05$, and
$\delta = 0.1$, respectively. 
} 
\label{scaldlca} 
\end{figure} 

\narrowtext
\begin{figure}
\includegraphics[width=0.95\linewidth]{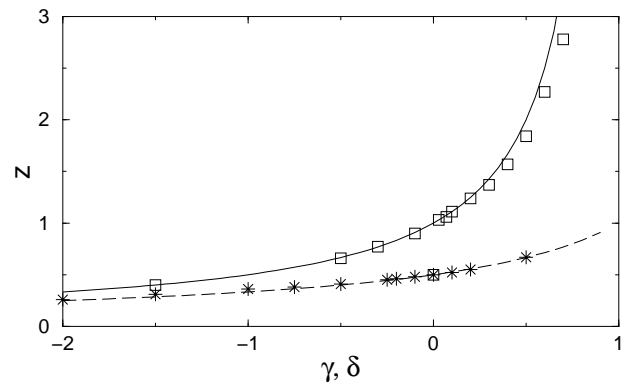}
\caption{Dynamic exponent from simulations either as a function of the
diffusion exponent $\gamma$~($*$) or the field exponent
$\delta$~($\square$). The solid
line is given by $1/(1-\delta)$ and the dashed one by $1/(2-\gamma)$.}
\label{zdelta} 
\end{figure}

We first consider purely diffusive dynamics, {\it i.e.},
$\Cf=0$. We obtain an excellent scaling for the cluster size distribution
using the scaling form $n_s(t) = \avecs(t)^{-2}f(s/\avecs(t))$
(Fig.~\ref{scaldlca}) and the known~\cite{Kang:PRA33,Miyazima:PRA36}
result for the dynamic exponent $z = 1/(2-\gamma)$ (Fig.~\ref{zdelta}). 

\narrowtext
\begin{figure}
\includegraphics[width=0.95\linewidth]{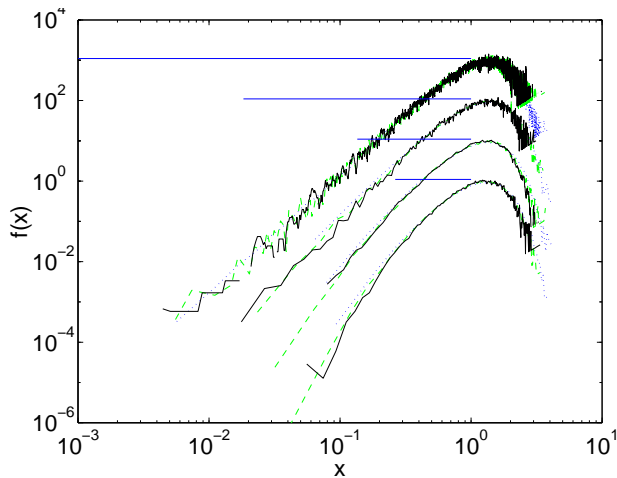}
\caption{
The scaling functions as a function of the scaling variable
for DLCA for $\gamma = -0.05$, $-0.25$, $-0.50$, $-0.75$ (from top to
bottom) at the times $10^4$~($\cdots$), 
$10^5$(-- --), and $9 \cdot 10^5$~(---). System size $L=5\times10^5$
and data are averaged over $25$ runs except for $\gamma =-0.05$ over
491 runs. Horizontal lines show the
crossover region $\exp(1/\gamma) \le x \le 1$ where the scaling
functions show typical class~II behavior. The data for various
$\gamma$-values have been shifted
in the vertical direction to make the figure clearer.} 
\label{crossover} 
\end{figure}

The decay of the scaling function near $x=0$ 
depends on the sign of $\gamma$ and there is a transition from
class~III ($\gamma<\gamma_c$) to class~II ($\gamma \ge \gamma_c$) at
$\gamma_c=0$ in accordance with the mean-field analysis. 
However, the transition between the algebraic and non-algebraic decay
of the scaling function is plagued by
strong crossover effects. This is illustrated in
figure~\ref{crossover} where the scaling functions are presented for
several values of the diffusion exponent. The crossover behavior is in
excellent agreement with the mean-field theory according to which the
kernels in classes~I and III show typical class~II behavior for
intermediate $x$ values: $\exp(-1/|\mu|) \ll x \ll
1$~\cite{vanDongen:PRL54}. In our case the $\mu = \gamma$ and the 
intermediate $x$ region is presented by horizontal lines in
figure~\ref{crossover}. The dynamics for $\gamma = 0$ can be solved
exactly to establish that the DLCA belongs
to class~II at $\gamma_c$. The exact result for the cluster size
distribution is $n_s(t) =
\exp(-\xi)[I_{s-1}(\xi)-I_{s+1}(\xi)]$, where $T=4 D_1 t$ and
$I_s(T)$ is the modified Bessel function~\cite{Spouge:PRL60}. This
gives $f(x) \simeq x \exp(-C x^2) \sim x$ ($x\to 0$), where the
constant $C$ depends on 
the average used to calculate the mean cluster size.  

As the scaling function decays faster than a power-law in class~III
the polydispersity exponent $\tau$ is well-defined only for $\gamma
\ge 0$. Although the statistics is insufficient for
a direct determination of the relationship $\tau(\gamma)$ the fits
to the scaling function show that $\tau$ 
increases monotonically with increasing $\gamma$
so that $\tau = 0$ at about $\gamma \approx 0.7$. 
The scaling theory states that for class II
$n_s(t) \sim s^{-\tau}t^{-w}$ for $1 \ll s \ll \avecs$ and $t \to
\infty$ with the scaling relation $w=(2-\tau)
z$~\cite{vanDongen:PRL54}. The exponent $w$ can be 
obtained more accurately from simulations than $\tau$. A careful
analysis of the data shows that $w$ is roughly a constant
$w \approx 1.50 \pm 0.05$ for $\gamma \in [0,0.5]$. However, 
$w$ cannot be independent of $\gamma$ since necessarily $w \ge z$ which
diverges when $\gamma \to 2$. By approximating $w\approx1.50$ near
$\gamma=0$ leads to $\tau(\gamma) \approx 1.50\gamma-1.00$, which is
zero at $\gamma_0 \approx 0.67$ (compare with the actual result above).
This approximation is consistent with the exact value $\tau(0) =
-1$~\cite{Spouge:PRL60}.  

Note that the point $\gamma_0 \approx 0.7$ at 
which the cluster size distribution changes from a non-monotonic function
to a monotonic one is not the same as the transition point between
the classes $\gamma_c=0$. In the literature it has been argued that in
two and three 
dimensions $\gamma_c$ is negative but these arguments rely on
the fact the cluster size distribution would change to a non-monotonic
function at the same point~\cite{Meakin:PRB31}. As this is clearly 
not the case in one dimension it is highly probable that 
$\gamma_c=0$ in higher dimensions, too. 

The corresponding FDCA-simulations are done using $\Cpar=0$. 
Figure~\ref{zdelta} shows also for this case the dynamic 
exponent as a function of the field exponent together with the
mean-field prediction. 
The agreement is excellent except for
$\delta>0.3$ for which values the asymptotic regime has not been
reached.  $\delta=0$ is a special point: all the clusters move with
the same velocity but the algorithm itself causes intrinsic diffusion 
resulting in the standard random walk value $z(\delta=0)=1/2$.

As in the purely diffusive case the cluster size distribution exhibits
scale invariance $n_s(t) = \avecs(t)^{-2}g(s/\avecs(t))$ but now with a
bell-shaped scaling function 
$g(x) \sim \exp(-x^{-|\mu|})$ as $x \to 0$ (see figure~\ref{scaldlca}).
Thus FDCA belongs to class~III. No indication of belonging
to the class II is seen in the range $-1.5 \le \delta \le 0.7$ in
contradiction with the mean-field theory. 
The absence of the transition shows that although the mean-field
analysis gives the correct dynamic exponent it fails in the case of
the scaling function.
This is not surprising since the spatial fluctuations expected to be
important in low dimensions are completely neglected in
equation~\eqref{Smoluchowski}. 
Furthermore, for $\delta >0$ the probability for collisions of large 
clusters with large ones is relatively small compared to large-small
collisions since the decisive factor is the velocity difference,
not the high mobility of large ones.

The case $\delta=\gamma=0$ of FDCA is interestingly enough related to a
driven diffusive Ising system (DDS). The low temperature coarsening in an Ising
chain with conserved magnetization and subject to a small external
force can be mapped almost exactly to the diffusion of domains with a
size-independent diffusion constant~\cite{Cornell:PRE54}. The fact
that the mapping is not quite one-to-one is reflected in the behavior
of dimers in the DDS. They perform long-range hopping which results
in another characteristic length scale in the problem~\cite{Spirin:PRE60}. 
As a consequence the domain length distribution does not obey the usual
dynamic scaling for small cluster sizes like it does in the FDCA,
although the domain size distributions are otherwise practically the 
same~\cite{Spirin:PRE60}. 

\narrowtext
\begin{figure}
\centering
\includegraphics[angle=0,width=0.95 \linewidth]{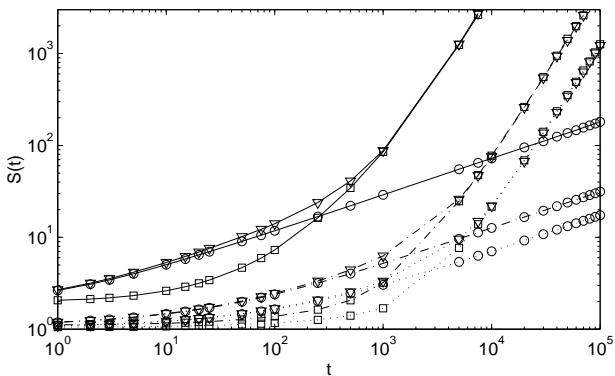}
\caption{Average cluster size for various mobilities and
concentrations for $\gamma = -0.5$ and $\delta = 0.5$ in the diffusive
$\Cpar = 1$, $\Cf=0$~($\bigcirc$), driven $\Cpar = 0$, $\Cf
= 0.05$~($\square$), and driven diffusive 
$\Cpar = 1$, $\Cf = 0.05$~($\nabla$) cases. 
Data are averaged over 50 runs and system sizes are $10^6$,
$5\times10^5$, and $10^5$ for concentrations
$\phi = 0.05$~($\cdots$), 0.1~($-\cdot$), and 0.5~(---), respectively.}
\label{diffusion_and_field} 
\end{figure}

Figure~\ref{diffusion_and_field} shows the crossover from the
diffusion dominated growth to the field dominated one for three
different concentrations. Estimating the unknown parameter $A$ in
equation~\eqref{Scross1d} using the scaling function of the diffusion
limited aggregation for $\gamma = -0.5$ gives $A
\approx0.2$. Equation~\eqref{Scross1d} gives the crossover sizes
$3, 4$, and $10$ for
concentrations $\phi = 0.05, 0.1$, and $0.5$, respectively. These
values agree reasonably well with the simulations as can be seen from
figure~\ref{diffusion_and_field}.

\narrowtext
\begin{figure}
\centering
\includegraphics[angle=-90,width=0.95\linewidth]{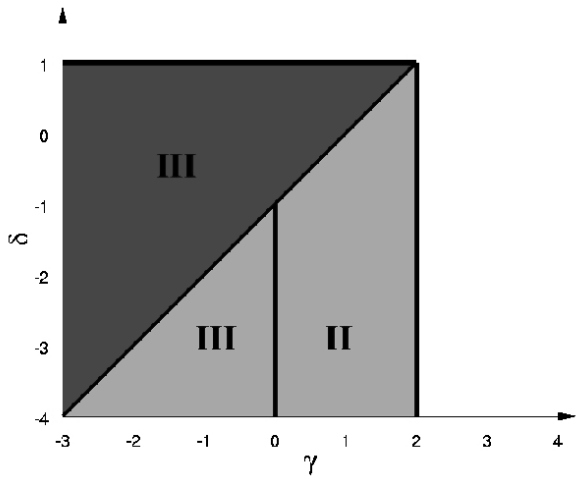}
\caption{The phase diagram in one dimension. Roman numbers
indicate the class of 
the aggregation process. Aggregation is dominated by the
diffusion~(light gray), the 
field~(dark gray) or by a gelation transition~(white).
} \label{phasediag}
\end{figure}

\section{Discussion} \label{dissec}

The results of our study are summarized in figure~\ref{phasediag}
which shows the dynamic 
phase diagram with four different regions. The aggregation is 
dominated by the field or the diffusion.
At the phase boundary $\delta = \gamma-1$ the two processes give the
same dynamic exponent. It is unclear
which one of the aggregation mechanisms determines
the asymptotic scaling behavior at the boundary.
The diffusive phase is split into two subphases according to the
dominating aggregation mechanism. The dynamics 
may also be so fast that the systems gels in a finite time. 

Although this paper has considered $d=1$ we can also discuss
the $d>1$ case. Here, complications arise because the clusters may
have a fractal structure. For the field-driven  
aggregation the clusters will in any case become anisotropic with a 
preferred orientation in the field direction. 
We believe both of these complications affect only the
phase boundaries of the dynamic phase diagram but leave its general
structure invariant if temporal scaling can be assumed. One particular
issue is the existence of a field-dominated phase with the scaling
function belonging to class II. The comparison
of the mean-field approach and simulations in higher dimensions is
left for a forthcoming study. The exact location of the phase
boundaries would be an interesting problem also when it comes
to applications to experiments. 

In conclusion, we have studied one-dimensional driven diffusive
cluster--cluster aggregation. We have shown how the scaling function 
depends on the cluster mobilities with diffusive or ballistic dynamics,
or both. For the field dominated case the dynamic
exponent can be obtained from simple mean-field calculations, which
together with the simulation results may be used to obtain the phase
boundaries in the dynamic phase diagram. This shows four different
phases in the aggregation depending on the relative strength
of the diffusion and the field.

\end{multicols}

\end{document}